%% file: rehlers.HP2020.tex
\documentclass[a4paper,11pt]{article}
\usepackage{pos}
\usepackage{tikz}
\usepackage{calc}
\usetikzlibrary{shapes,arrows}
\usetikzlibrary{fit,calc}
\usetikzlibrary{positioning}

\newcommand\AliceApprovedLabel[4][1.0]{%
    \begin{scope}[x={($ (#2.south east) - (#2.south west) $ )},y={( $ (#2.north west) - (#2.south west)$ )}, shift={(#2.south west)}]
        \node[anchor=south west, inner sep={0.005 * #1 * \textwidth}] at (0, 0) {\fontfamily{pcr}\selectfont\resizebox{\textwidth*\real{#3}*\real{#1}}{!}{\textcolor[rgb]{0.698,0.698,0.698}{\textbf{#4}}}};
    \end{scope}
}
\newcommand\AlicePreliminary[3][1.0]{%
    \node[anchor=south west,inner sep=0] (image) at (0,0) {#2};
    \AliceApprovedLabel[#1]{image}{0.185}{#3}
}
\newcommand\AlicePreliminaryStandalone[3][1.0]{%
  \begin{tikzpicture}
    \AlicePreliminary[#1]{#2}{#3}
  \end{tikzpicture}
}
\newcommand\AliceSimulation[3][1.0]{%
    \node[anchor=south west,inner sep=0] (image) at (0,0) {#2};
    \AliceApprovedLabel[#1]{image}{0.195}{#3}
}
\newcommand\AliceSimulationStandalone[3][1.0]{%
  \begin{tikzpicture}
    \AliceSimulation[#1]{#2}{#3}
  \end{tikzpicture}
}

\input{shared.tex}

\title{Investigating Hard Splittings via Jet Substructure in \pp{} and \PbPb{} Collisions at $\sqrts{} = 5.02$ TeV with ALICE}
\ShortTitle{Investigating Hard Splittings via Jet Substructure with ALICE}

\manuallySeparateAuthors{}
\author*{Raymond Ehlers}
\author{on behalf of the ALICE Collaboration}

\affiliation{Oak Ridge National Laboratory}

\emailAdd{raymond.ehlers@cern.ch}

\abstract{
Jets lose energy as they propagate through the Quark-Gluon Plasma, modifying their parton shower.
Jet substructure, which provides access to the evolution of jet splittings, is expected to be sensitive to interactions
between the medium and the jet, providing the opportunity to further constrain both jet and medium properties.
By utilizing grooming techniques, we can focus on the most pertinent hard splittings.
Of particular interest is the search for large transverse momentum kicks which may
indicate the presence of point-like scatters within the Quark-Gluon Plasma. We explore the jet substructure of inclusive jets
in \pp{} and \PbPb{} collisions at $\sqrts{} = 5.02$ TeV, utilizing Soft Drop and other
grooming methods, as well as the Lund Plane, in order to access the hardest jet splitting, with a particular
focus on the hardest $k_{\text{T}}$ splitting.
}

\FullConference{%
  HardProbes2020\\
  1-6 June 2020\\
  Austin, Texas}

\begin{document}
\maketitle

\input{latex/introduction.tex}
\input{latex/PbPbSubstructure.tex}
\input{latex/hardestKt.tex}
\input{latex/conclusions.tex}

\scriptsize
\bibliographystyle{style/JHEP}
\bibliography{rehlers.HP2020.bib}

\end{document}

%% file: shared.tex
\usepackage{amsmath}

\newcommand{\pp}{\ensuremath{\text{p\kern-0.05em p}}}

\newcommand{\PbPb}{\ensuremath{\mbox{Pb--Pb}}}

\newcommand{\sqrts}{\ensuremath{\sqrt{s_{\text{NN}}}}}

\newcommand{\figRef}[1]{Fig.~\ref{#1}}

\newcommand{\GeVc}{\ensuremath{\text{GeV}\kern-0.05em/\kern-0.02em c}}

\newcommand{\pT}{\ensuremath{p_{\text{T}}}}

\newcommand{\zg}{\ensuremath{z_{\text{g}}}}
\newcommand{\zcut}{\ensuremath{z_{\text{cut}}}}
\newcommand{\Rg}{\ensuremath{R_{\text{g}}}}
\newcommand{\nsd}{\ensuremath{n_{\text{SD}}}}
\newcommand{\kT}{\ensuremath{k_{\text{T}}}}
\newcommand{\deltaR}{\ensuremath{\Delta R}}



%% file: latex/introduction.tex
\hypertarget{introduction}{%
\section{Introduction}\label{introduction}}

As partons from high momentum transfer processes propagate through the
medium, they interact with it, losing energy and modifying their parton
shower. These interactions between the jet and the hot and dense QCD
medium known as the Quark-Gluon Plasma (QGP) are expected to modify the
internal jet structure. Jet substructure measurements provide access to
jet splittings, and consequently may be sensitive to these
modifications.

To perform such measurements, selections are often made on the jet
splitting properties via grooming techniques
\cite{Larkoski:2014wba,Mehtar-Tani:2019rrk,Acharya:2020aa}. In \pp{}
collisions, grooming limits contamination of the jet shower by soft QCD
processes, while in \PbPb{} collisions, grooming helps select the hard
component of quenched jets. Utilizing these techniques, substructure may
provide direct access to medium properties such as color coherence
\cite{Casalderrey-Solana:2020aa}, or quasi-particle structure which can
be searched for indirectly by looking for large angle Moliere scattering
\cite{DEramo:2013aa}.

ALICE \cite{Abelev:2014ffa} is well suited for performing jet
substructure measurements due to the precision tracking provided by the
Inner Tracking System and Time Projection Chamber in the central barrel.
For these measurements, charged-particle \(R=0.4\) anti-\(\kT{}\) jets
were reconstructed using FastJet 3.2.1 \cite{Cacciari:2011ma} in \pp{}
and \PbPb{} collisions at \(\sqrts{}=5.02\) TeV that were collected in
2017 and 2018 respectively. Jets were required to be contained entirely
within the ALICE central barrel acceptance. In \PbPb{} collisions,
background subtraction is of particular importance. For substructure
analysis, ALICE performs background subtraction via Constituent
Subtraction \cite{Berta:2019aa}. Performance was optimized with the goal
of reducing the background contribution while minimizing any possible
bias on the substructure variables. These studies determined an optimal
value of \(\Delta R^{\text{max}} = 0.6\).

%% file: latex/PbPbSubstructure.tex
\hypertarget{groomed-jet-substructure-in-3050-pbpb-collisions}{%
\section{\texorpdfstring{Groomed Jet Substructure in 30--50\%
\(\PbPb{}\)
Collisions}{Groomed Jet Substructure in 30--50\% \textbackslash PbPb\{\} Collisions}}\label{groomed-jet-substructure-in-3050-pbpb-collisions}}

To characterize jet substructure in 30--50\% semi-central \PbPb{}
collisions, the Soft Drop grooming algorithm \cite{Larkoski:2014wba} was
utilized to select the first sufficiently hard splitting. In particular,
we measured the shared momentum fraction, \zg{}, the angular separation
between the subjets from the selected splitting, \Rg{}, and the number
of splittings until finding the hard splitting, \nsd{}
\cite{Larkoski:2014wba,Acharya:2020aa}. In order to avoid background
contaminated splittings, splittings were considered sufficiently hard
when they passed the requirement of \(\zcut{} = 0.2\) or
\(\zcut{} = 0.4\). For each variable, Bayesian iterative 2D unfolding
was utilized to correct for background fluctuations and detector effects
\cite{Adye:1349242}.

\begin{figure}[t]
    \centering
    \AlicePreliminaryStandalone[0.425]{\includegraphics[width=0.4\textwidth]{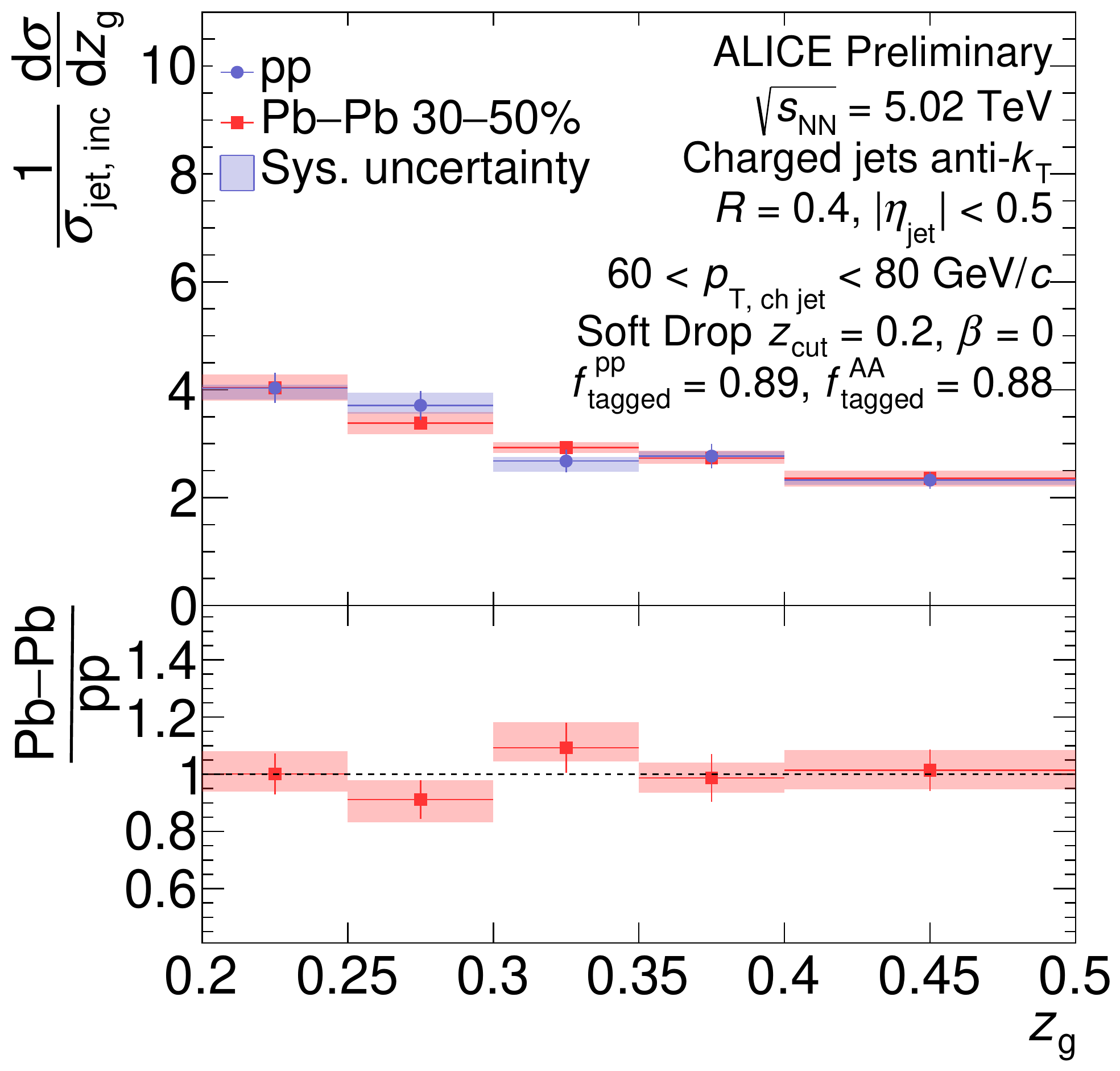}}{ALI-PREL-352054}
    \AlicePreliminaryStandalone[0.425]{\includegraphics[width=0.4\textwidth]{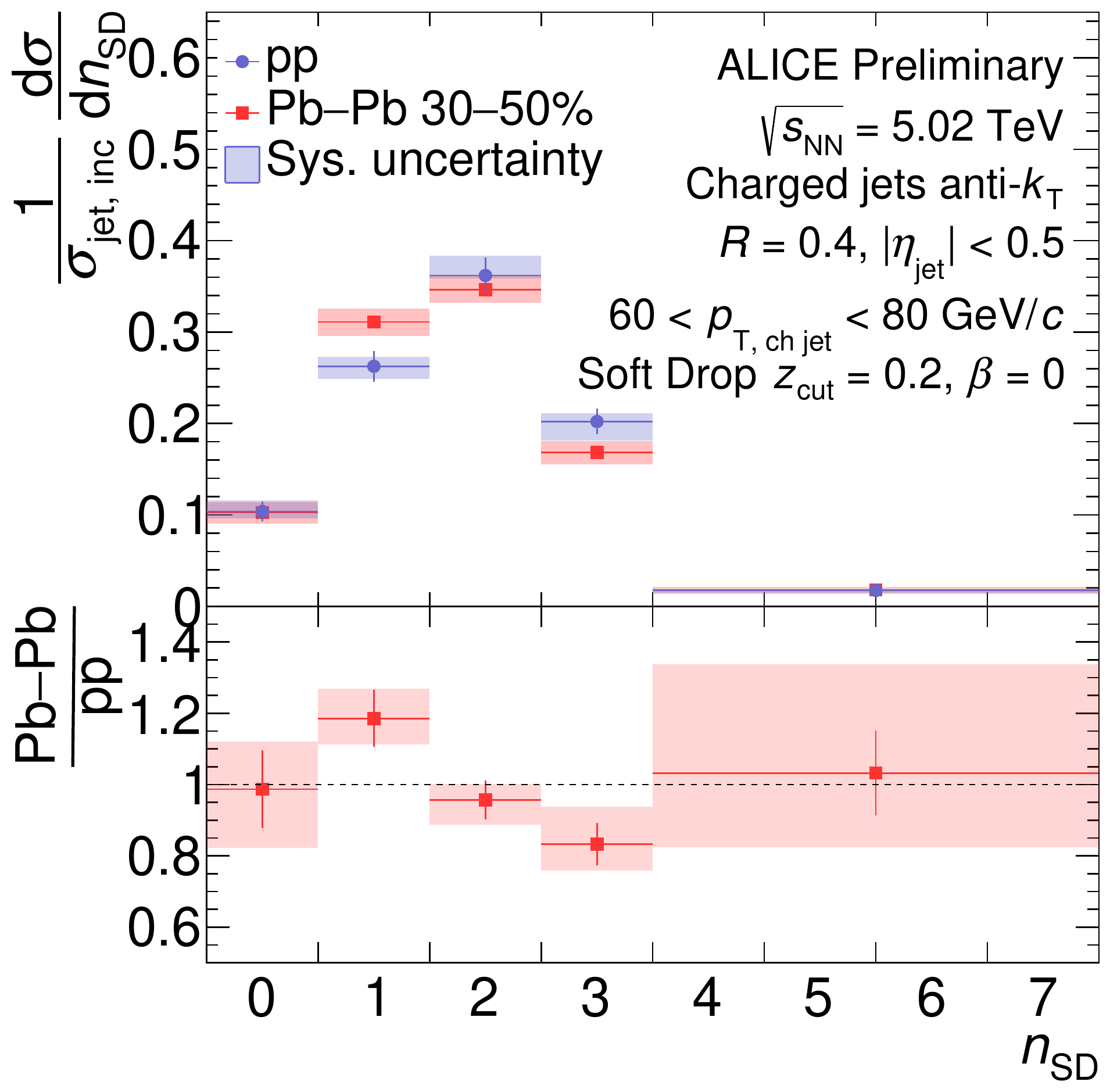}}{ALI-PREL-352059}
    \caption{Measurement of $\zg{}$ (left) and $\nsd{}$ (right) for $R=0.4$ charged-particle jets in $60 < \pT{}_{\text{ch,jet}} < 80$ \GeVc{} in \pp{} and \PbPb{} collisions. Both measurements are consistent with no modification within experimental uncertainties.}
    \label{fig:PbPbSemiCentralZgNsd}
\end{figure}

\begin{figure}[b]
    \centering
    \AlicePreliminaryStandalone[0.425]{\includegraphics[width=0.4\textwidth]{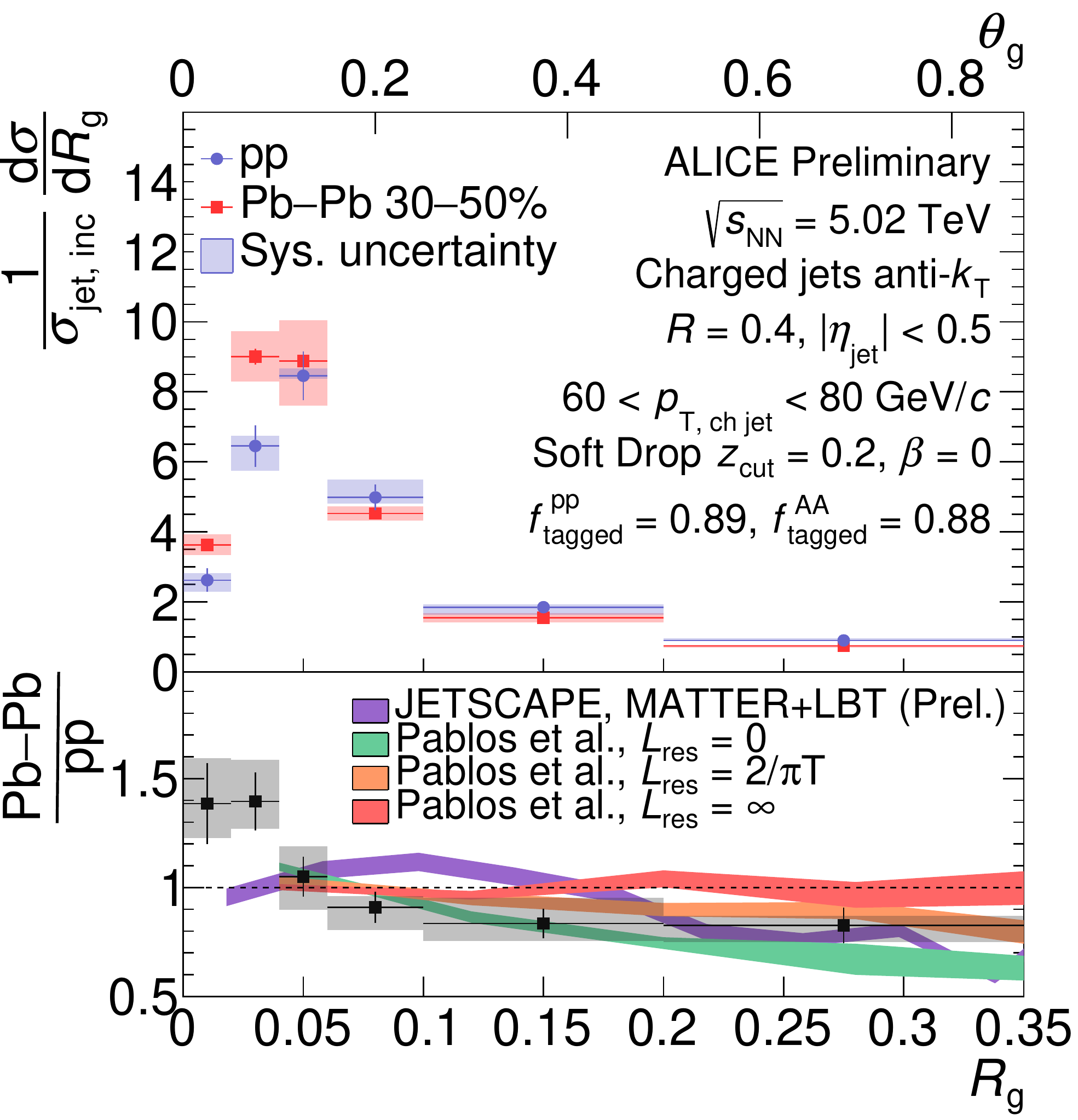}}{ALI-PREL-353300}
    \AlicePreliminaryStandalone[0.425]{\includegraphics[width=0.4\textwidth]{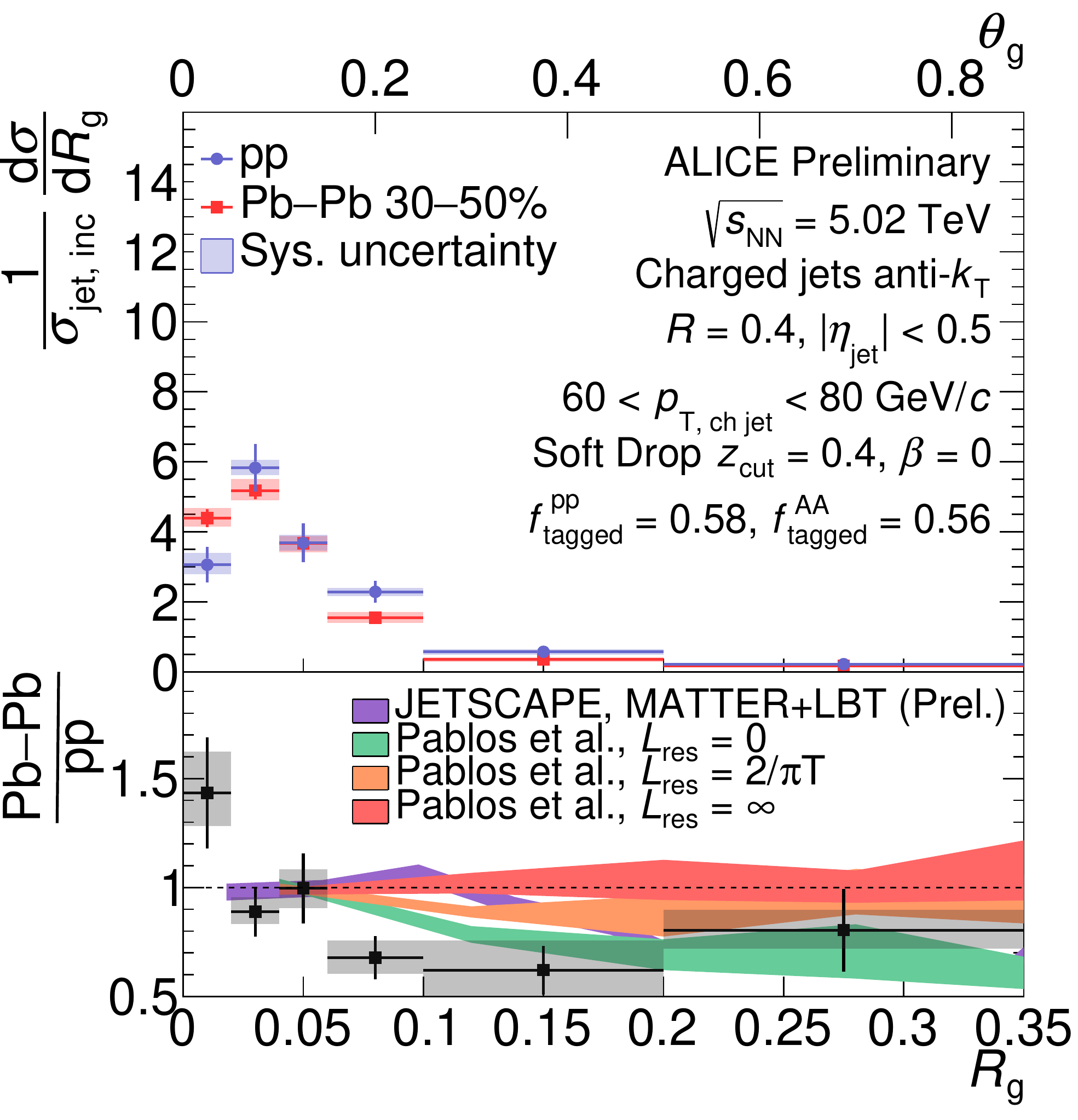}}{ALI-PREL-353397}
    \caption{Measurement of $\Rg{}$ for $\zcut{} = 0.2$ (left) and $\zcut{} = 0.4$ (right)  for $R=0.4$ charged-particle jets in $60 < \pT{}_{\text{ch,jet}} < 80$ \GeVc{} in \pp{} and \PbPb{} collisions. Both values of $\zcut{}$ show enhancement for small angle splittings, as well as suppression for large angle splittings. The models are described in the text.}
    \label{fig:PbPbSemiCentralRg}
\end{figure}

The results of this analysis for jets measured within
\(60 < \pT{}_{\text{ch,jet}} < 80\) \GeVc{} are shown in
\figRef{fig:PbPbSemiCentralZgNsd} and \figRef{fig:PbPbSemiCentralRg}. In
the left panel of \figRef{fig:PbPbSemiCentralZgNsd}, \(\zg{}\) measured
in \PbPb{} collisions is compared to the same measurement in \pp{}
collisions for \(\zcut{} = 0.2\). Within experimental uncertainties,
\(\zg{}\) is consistent with no modification. \nsd{} is shown in the
right panel of \figRef{fig:PbPbSemiCentralZgNsd}, and is also consistent
with no modification relative to \pp{} collisions. The left and right
panels of \figRef{fig:PbPbSemiCentralRg} show \(\Rg{}\) measured in
\PbPb{} and \pp{} collisions for \(\zcut{}\) values of 0.2 and 0.4,
respectively. Both panels show similar behavior, with small angle
splittings enhanced in \PbPb{} collisions, while large angle splittings
are suppressed. The measurements were tested for consistency with no
modification of the ratio from unity by adding the statistical and
systematic uncertainties in quadrature. Within the context of this
simple metric, both measurements were found to be inconsistent with no
modification (\(p=0.03\)). The measurements are also compared against
model predictions from JETSCAPE \cite{Putschke:2019yrg}, and Pablos et
al \cite{Casalderrey-Solana:2020aa}. Given the experimental
uncertainties, these measurements may have the potential to provide
differentiation between model settings and insight into color coherence.

%% file: latex/hardestKt.tex
\hypertarget{hardest-kt-in-pp-and-pbpb-collisions}{%
\section{\texorpdfstring{Hardest \(\kT{}\) in \(\pp{}\) and \(\PbPb{}\)
Collisions}{Hardest \textbackslash kT\{\} in \textbackslash pp\{\} and \textbackslash PbPb\{\} Collisions}}\label{hardest-kt-in-pp-and-pbpb-collisions}}

Beyond measurements of the substructure variables themselves, can jet
substructure be used as a tool to isolate the effects of jet-medium
interactions? To address this question, we consider the search for the
presence of medium scattering centers via the measurement of rare, wide
angle scattering relative to the trigger jet axis, known as Moliere
Scattering \cite{DEramo:2013aa}. Searches by ALICE using large-angle
hadron-jet decorrelation at \(\sqrts{} = 2.76\) are consistent with no
medium-induced acoplanarity of recoil jets within measurement
uncertainties \cite{Adam:2015doa}. As an alternative, we investigate the
possibility of using jet substructure as a tool to search for these
medium scattering centers. As subjets propagate through the medium, they
may be deflected by a scattering center, which should lead to an
increase in the yield of high-\kT{} splittings in \PbPb{} collisions
relative to \pp{} collisions.

In order to identify the hardest \kT{} splitting, we investigated four
methods: leading \kT{}, leading \kT{} for all \(z > 0.2\) splittings,
and Dynamical Grooming \cite{Mehtar-Tani:2019rrk} with \(a=1\) (known as
\kT{}Drop) and \(a=2\) (known as timeDrop). The leading \kT{} selects
the maximum \kT{} splitting from all available splittings, while the
\(z > 0.2\) variation selects the maximum \kT{} out of all splittings
with \(z > 0.2\). Dynamical Grooming utilizes a hardness measure,
\(\kappa^{(a)} = z_{i} (1-z_{i})p_{\text{T,i}}(\frac{\deltaR{}}{R})^{a}\),
where \(i\) identifies a particular splitting, to determine the hardest
splitting \cite{Mehtar-Tani:2019rrk}. All methods consider all iterative
splittings, following the leading subjet to the next splitting.

\begin{figure}[t]
    \centering
    \AliceSimulationStandalone[0.425]{\includegraphics[width=0.4\textwidth]{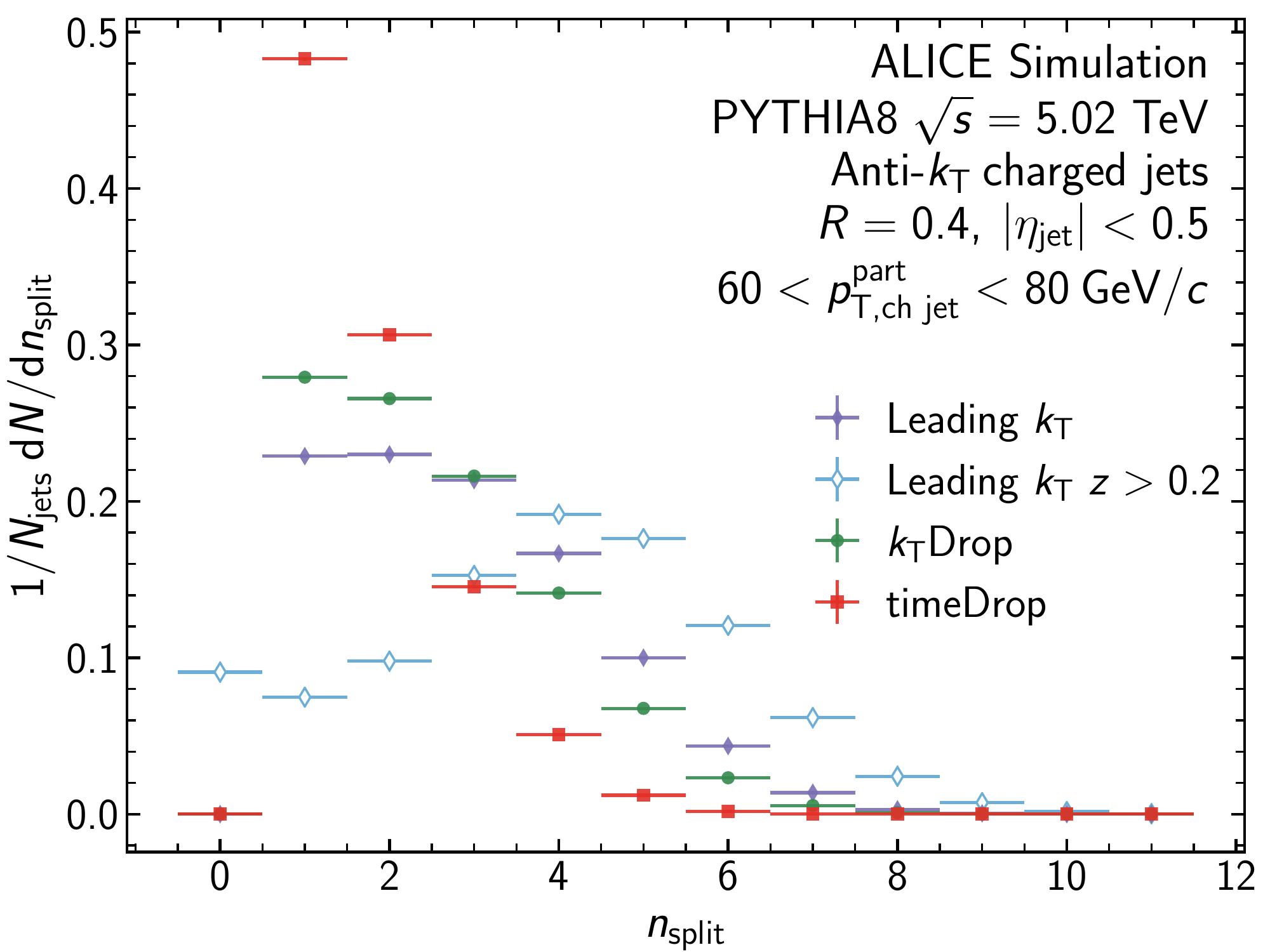}}{ALI-SIMUL-352437}
    \AliceSimulationStandalone[0.425]{\includegraphics[width=0.4\textwidth]{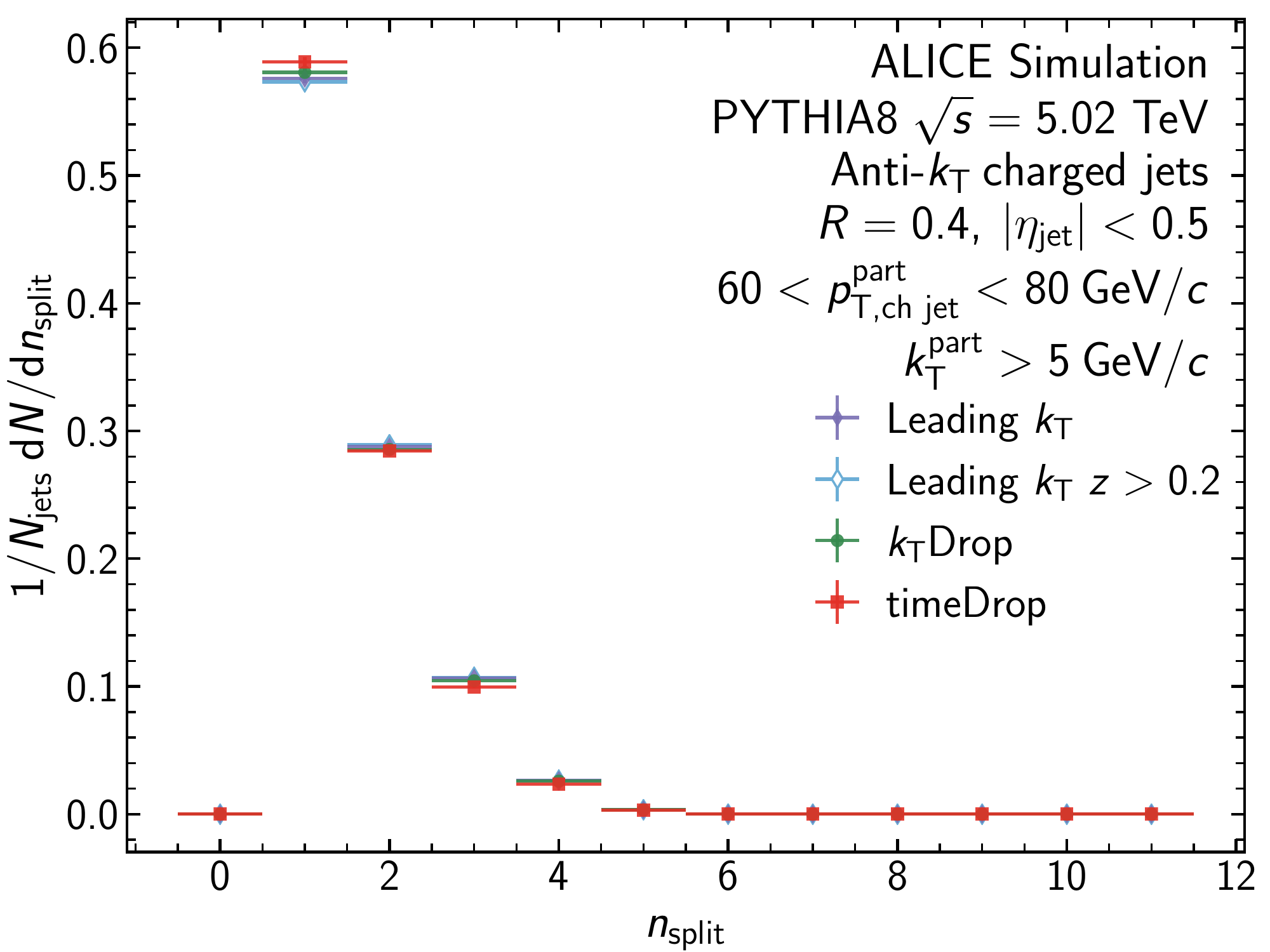}}{ALI-SIMUL-352442}
    \caption{Measurement of the number of splittings until the hardest splitting is identified $n_{\text{split}}$ for inclusive \kT{} (left) and $\kT{} > 5$ \GeVc{} (right) for $R=0.4$ charged-particle jets in $60 < \pT{}_{\text{ch,jet}} < 80$ \GeVc{} in PYTHIA 8 Monash 2013. The splittings selected by the different grooming methods converge at high-$\kT{}$.}
    \label{fig:ktNumberOfSplittings}
\end{figure}

To initially study these grooming methods, they were applied to PYTHIA 8
Monash 2013 \cite{Sjostrand:2015aa} at particle level. The performance
was characterized through properties such as the number of splittings
until the hardest splitting is identified, \(n_{\text{split}}\), as
shown in \figRef{fig:ktNumberOfSplittings}. These studies demonstrated
that although the grooming methods perform differently at low \(\kT{}\),
for sufficiently high \(\kT{}\) splittings (here, \(\kT{} > 5\)
\GeVc{}), all grooming methods converge, selecting the same splittings.

\begin{figure}[th]
    \centering
    \AlicePreliminaryStandalone[0.7]{\includegraphics[width=0.7\textwidth]{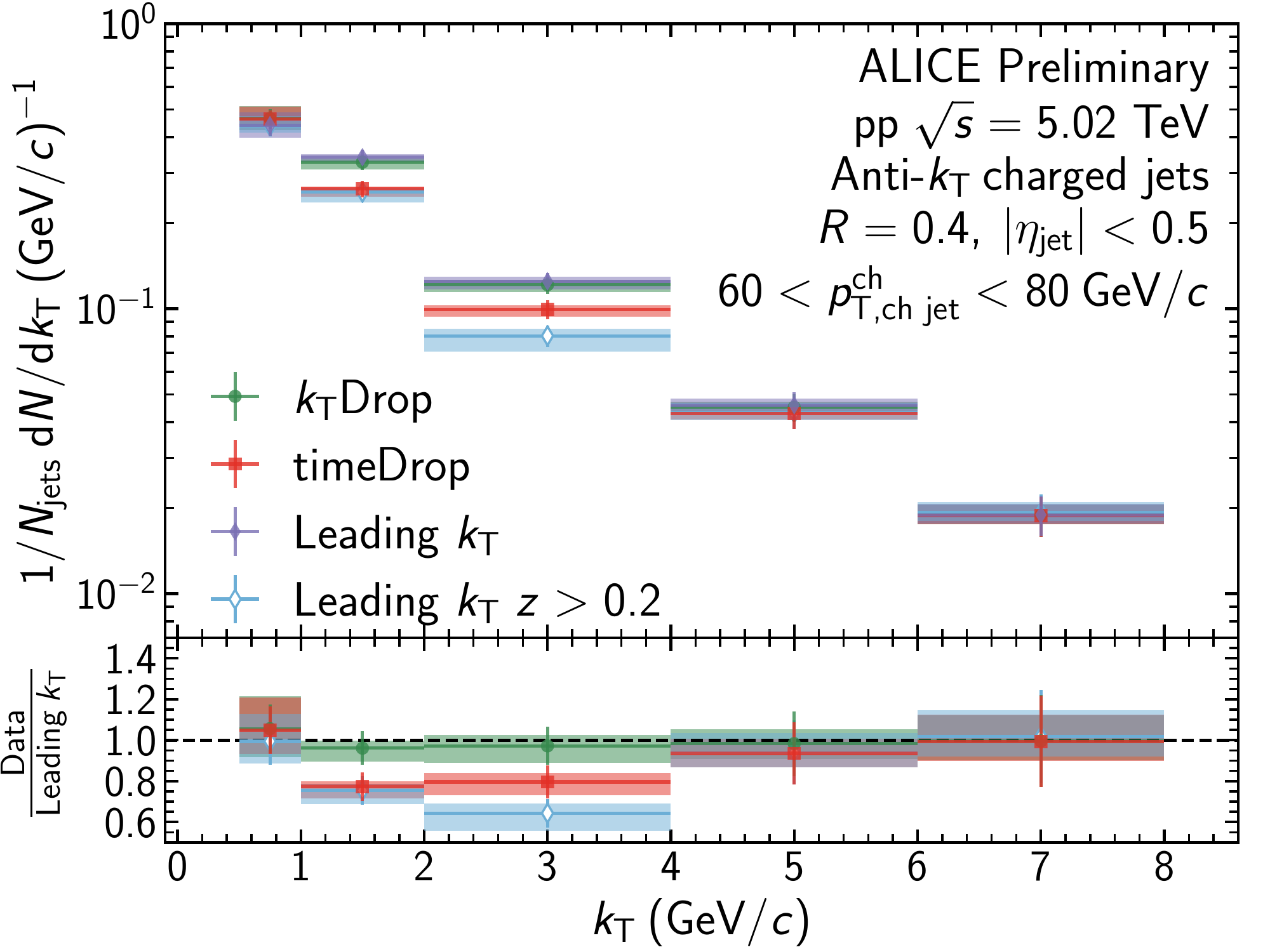}}{ALI-PREL-352215}
    \caption{Measurement of the hardest \kT{} splitting for four grooming methods for $R=0.4$ charged-particle jets in $60 < \pT{}_{\text{ch,jet}} < 80$ \GeVc{} in \pp{} collisions and PYTHIA 8 Monash 2013. The splittings selected by the different grooming methods converge at high-$\kT{}$. PYTHIA is broadly consistent with the data.}
    \label{fig:ktGroomingComparison}
\end{figure}

With these comparisons in mind, the four grooming methods were applied
to measure the hardest \(\kT{}\) in \pp{} collisions at
\(\sqrt{s} = 5.02\) TeV, as shown in \figRef{fig:ktGroomingComparison}.
Bayesian iterative 2D unfolding was again employed. For \(\kT{} > 4\)
\GeVc{} splittings, the \(\kT{}\) spectra converge for all of the
grooming methods. This behavior is consistent with the earlier PYTHIA
studies. Each measurement was also directly compared to PYTHIA 8 Monash
2013 by applying the same grooming methods. PYTHIA is broadly consistent
with the data within the statistical and systematic uncertainties,
although there is a hint of a shape difference that is consistent
between all grooming methods.

In order to assess the prospects for measuring the hardest \kT{} in
\PbPb{} collisions, we studied the correlation between the hardest \kT{}
in the PYTHIA splitting graph vs that which is found via declustering
\(R=0.8\) jets. Previous studies showed a clear correlation between the
graph and the declustering splittings \cite{EMMI}. To study this
correlation in a large background environment, the PYTHIA particles were
embedded into a thermal background tuned to 0--10\% central data. Using
this thermal model, the background contribution is apparent at low to
intermediate \(\kT{}\), but a strong correlation is observed at
high-\(\kT{}\), encouraging the possibility of such a measurement in
\PbPb{}.

%% file: latex/conclusions.tex
\hypertarget{conclusions-and-outlook}{%
\section{Conclusions and Outlook}\label{conclusions-and-outlook}}

We presented fully unfolded \(\zg{}\), \(\nsd{}\), and \(\Rg{}\)
measurements in 30--50\% semi-central \(\PbPb{}\) and \(\pp{}\)
collisions at \(\sqrts{} = 5.02\) TeV. \(\zg{}\) and \(\nsd{}\) are
consistent with no modification in \PbPb{} collisions relative to \pp{}
collisions, while \(\Rg{}\) shows enhancement for small angle splittings
and suppression for large angle splittings. These modifications are
consistent for both \(\zcut{} = 0.2\) and 0.4. The hardest \(\kT{}\)
splittings were measured in \(\pp{}\) collisions for a variety of
grooming methods. The grooming methods selected a consistent set of
splittings for \(\kT{} > 4\) \GeVc{}. The prospects for measuring the
hardest \(\kT{}\) splittings in \PbPb{} were also explored as a step
towards applying jet substructure as tool to search for point-like
scattering centers in the medium via large angle scattering.